\documentclass{emulateapj}
\slugcomment{{\sc \bf Accepted to ApJL --- Preprint Version }} 
\usepackage{graphicx,natbib}

\shorttitle{AB Aur's Disk in Polarized Light}
\shortauthors{Perrin et al.}

\begin{document}

\singlespace

\title{The Case of AB Aurigae's Disk in Polarized Light: \\
Is there truly a gap?
}
\author{Marshall D. Perrin\altaffilmark{1}}
\affil{Division of Astronomy, University
of California, Los Angeles, CA 90095\\
mperrin@ucla.edu}
\author{Glenn Schneider}
\affil{Steward Observatory, The University of Arizona, 933 North Cherry
Avenue, Tucson, AZ 85721}
\author{Gaspard Duchene}
\affil{University of California, Berkeley, Berkeley CA 94720, and\\
Universit\'e Joseph Fourier - Grenoble 1 / CNRS,  Laboratoire
d'Astrophysique de Grenoble (LAOG)\\
UMR 5571, BP 53, 38041 Grenoble Cedex 09, France}
\author{Christophe Pinte}
\affil{
Universit\'e Joseph Fourier - Grenoble 1 / CNRS,  Laboratoire
d'Astrophysique de Grenoble (LAOG)\\
UMR 5571, BP 53, 38041 Grenoble Cedex 09, France}
\author{Carol A. Grady}
\affil{Eureka Scientific, 2452 Delmer, Suite 100, Oakland, CA
96002.;\\
Exoplanets and Stellar Astrophysics Laboratory, Code 667, NASA's
Goddard Space Flight Center, Greenbelt, MD 20771}
\author{John P. Wisniewski}
\affil{Department of Astronomy, University of Washington\\
Seattle, WA 98195-1580}
\and
\author{Dean C. Hines}
\affil{Space Science Institute, 4750 Walnut Street, Suite 205, Boulder,
CO 80301}

\altaffiltext{1}{Center for Adaptive Optics, 
University of California, Santa Cruz, CA 95064, U.S.A. }

\begin{abstract}
Using the NICMOS coronagraph, we have obtained high-contrast 2.0~\micron\ imaging
polarimetry and 1.1~\micron\ imaging of the circumstellar disk around AB Aurigae on angular scales of
0.3--3\arcsec\ (40--550~AU). Unlike previous observations, these data resolve the disk in both total and polarized intensity, allowing accurate measurement of the spatial variation of polarization fraction across the disk.
Using these observations we investigate the
apparent ``gap'' in the disk reported by Oppenheimer et
al. 2008.  In polarized intensity, the
NICMOS data closely reproduces the morphology seen by
Oppenheimer et al., yet in total intensity we find 
no evidence for a gap in either our 1.1 or 2.0~\micron\ 
images. We find instead that region has lower polarization fraction,
without a significant decrease in total scattered light, consistent with
expectations for back-scattered light on the far
side of an inclined disk. Radiative transfer models demonstrate
this explanation fits the observations. Geometrical scattering effects are entirely
sufficient to explain the observed morphology without any need to
invoke a gap or protoplanet at that location.
\end{abstract}


\section{Introduction}

AB Aurigae is one of the most intensively
studied of all Herbig Ae/Be stars, on account of its proximity,
brightness, and youth (distance $d=144$ pc; visual magnitude $V=7.04$ mag; age $<3$ Myr; spectral type A0e). 
Steadily improving observational
capabilities have yielded increasingly detailed views of its complex
and dusty environment, providing many insights into the nature of circumstellar disks
\citep[e.g.][]{Grady:1999p2806,Fukagawa:2004p87,Pietu:2005p1365}.

In particular, \citet[][hereafter Opp08]{Oppenheimer:2008p2388}
recently presented high-angular resolution, high contrast imaging 
polarimetry of AB Aur at 1.6 ${\mu}m$, obtained with the Lyot Project
coronagraph on the AEOS 3.6 m telescope.  These observations resolved 
the disk in polarized scattered light as close as 40 AU 
(0.3$''$) to the star.
In polarized light the disk is not axisymmetric, but instead shows an apparent gap or depleted region at a radius of $\sim100$ AU. 
Such a gap may be created by dynamical perturbations from
forming planets \citep[e.g.][]{2003ApJ...588.1110K,Wyatt:2005A&A...440..937W,JangCondell:2009p2820}.
Intriguingly, Opp08 report a faint point source within the gap, which they conjecture could be the perturbing 
object---though they are cautious with this identification due to its low statistical significance (2.8~$\sigma$).
The formation mechanism(s) of massive planets at large separations
remain highly uncertain \citep{2009arXiv0909.2662D,Nero:2009p2842};
any prospect for observing a planet forming \textit{in situ}
around AB Aur should be pursued to clarify this puzzle. 

\begin{figure}
  \includegraphics[width=3.5in]{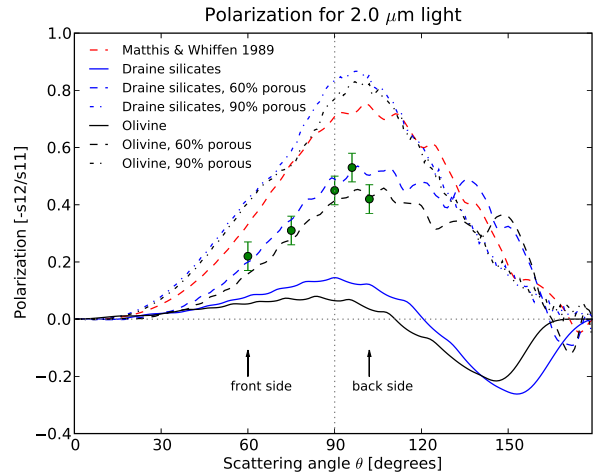}
  \caption{The induced polarization for some possible grain compositions, 
  demonstrating that polarization depends on both scattering angle and dust grain properties, 
  particularly composition and porosity.  
  Positive values indicate typical centrosymmetric linear
  polarization, while negative formal polarization denotes linear polarization 
  oriented radially.
  Polarization is maximized for porous
  grains, and for scattering angles slightly above 90$\degr$.  These models assume a power-law
  distribution of grain sizes from 0.03-200 $\micron$ with slope -3.5,
  typical for YSO disks. 
  See section 5.2 of \citet{Pinte:2008p2580} for a discussion of dust model
  and computation details.  The overplotted points show
  the fractional polarization observed around AB Aur; the inferred scattering
  angles are not symmetric due to the flaring of the disk surface, estimated $\sim10\degr$. Of the models
  shown, the 60\% porosity silicates provide the best fit.
  \label{scatteringprops}
  }
\end{figure}

One challenge in interpreting the data from Opp08 is that the disk is visible only in polarized
intensity, $P=\sqrt{Q^2+U^2}$ (where $Q$ and $U$ are the usual Stokes parameters; see
\citealt{Tinbergen1996} for a review of polarization fundamentals and
notation). In their total intensity image (Stokes $I$), the residual speckle halo of
the stellar PSF completely drowns out the disk's fainter light.  
The origin of features seen only in polarized light is ambiguous:
any observed spatial variation may be due either to variation in the \textit{total} amount of scattered light, or to a
change in the \textit{polarization fraction} of that light.  The
polarization induced by dust scattering depends strongly
on the scattering angle (see Figure \ref{scatteringprops}), allowing 
disk geometry or viewing angle to cause variations in the
observed polarization which might be mistaken for intrinsic substructure within
the disk.

In particular, for AB Aur the observed celestial position angle of the depleted region, $333\pm2\degr$
(Opp08) is precisely aligned with the disk's apparent
rotation axis as inferred from CO emission line kinematics (330--333$\degr$; \citealp{Corder:2005p2693,Pietu:2005p1365}). Is this alignment coincidental?

The most direct way to answer this question is to obtain 
images with enough contrast to directly detect the disk in total
intensity, and then 
calculate the polarization fraction, $p = P/I$. 
Such observations are best obtained with
the Hubble Space Telescope, whose ACS and NICMOS coronagraphs both
have provided sufficiently high contrast to precisely and accurately measure the
polarization of disk-scattered light
\citep[][Hines and Schneider, in prep]{Hines:2000p332,2007lyot.confQ..22S,Graham:2007p495}. 

In this paper we present new NICMOS coronagraphic imaging and polarimetry of AB
Aur which clarifies the nature of the dark ``gap'' observed by Opp08.
These data were obtained as part of a coronagraphic
polarimetry survey of young stars across a range of masses and ages
which will be reported more fully in future works
\citep[see][ for a brief overview]{Perrin2009Subaru}.

\section{Observations and Data Reduction}

Our observation and reduction strategies follow the standard recommendations for NICMOS coronagraphy.
We observed AB Aur on 2007 Sep 14 and 2007 Dec 21 as part
of program HST/GO 11155, in two visits identical except for a 136$\degr$ difference in
roll angle.
After centering AB Aur behind the NIC2 coronagraph hole, 
nine 192~s exposures were taken using the 2.0~\micron\ POL*L linear
polarizers, cycling between the 0, 120, and
240\degr\ polarizers after each exposure, followed by one 512~s
exposure with F110W. The telescope was then shifted to move AB
Aur a few arcseconds away from the occulting spot, after which we took
two dithered 4 s exposures in both POL0L and F110W 
for photometry on the unocculted star. 

Achieving high contrast with NICMOS requires 
subtracting a color-matched point spread function (PSF), and for polarimetric
observations, the reference star must be
unpolarized to minimize systematic biases. 
(The direct stellar light from AB Aur should be unpolarized
or nearly so, due to the low line-of-sight extinction, $A_V=0.25$;
\citealp{Roberge:2001p2821})
It
proved challenging to identify PSF stars which are unpolarized yet
also 
sufficiently red to match AB Aur's color ($H-K$=0.832), 
which is redder than the Rayleigh-Jeans slope, hence 
anything so red must either have nonthermal emission (often
highly polarized) or else be dusty and extincted (likewise polarizing).
After some consideration we identified nearby M dwarfs as the best
candidate PSF references, and
therefore observed Proxima
Centauri, GJ 273, and GJ 447 one visit each using an identical observing strategy as
above. We supplemented our
program with additional PSF observations drawn from programs HST/GO 10847 and 10852.

Our data reduction approach follows that of 
Schneider et al.~2005. Briefly: starting with pipeline-reduced images from STScI, we
corrected for bad pixels and sky/thermal background emission in all
images, and then
median-combined the three coronagraphic images for each POL*L filter. 
The F110W coronagraphy and the unocculted imaging were reduced similarly.

Obtaining optimal PSF subtraction depends on accurate registration and flux-scaling. 
Starting from flux ratios derived from phometry of the unocculted
stars, we adopted two independent strategies to optimize the
subtractions: 
(1) A manual search visually compared subtractions
of different reference PSFs while interactively varying the
registration and scaling to minimize the residuals.
(2)
An automated algorithm evaluated subtractions across a grid in alignment parameter
space for each image pair. Each subtracted image was high-pass filtered
to reject diffuse light from the disk
while emphasizing features with angular scale comparable to the diffraction
limit (the characteristic size of speckle residuals). The variance on this angular scale was minimized to find the best subtraction.
The best subtractions were obtained using GJ 273 as the PSF; our two optimization approaches yielded subtracted datasets with polarization
fractions differing by $\lesssim6\%$, which we adopt as our polarization uncertainty.

Remaining artifacts in the subtracted images such as diffraction
spikes were masked out. The images were 
rectified for geometrical distortion, rotated
to a common orientation, and combined to produce final mosaics in each
filter.  From these images the Stokes parameters $I, Q$, and $U$ were
derived using the POLARIZE software \citep{Hines:2000p332}
which models the imperfect linear polarizers in NICMOS.  The output
polarized images were then smoothed by a one resolution element ($\sim$3 pixels $=0.22"$) Gaussian, but we present the
undersampled F110W data at full resolution.

The resulting images are shown in figure \ref{heximages}. The F110W
subtractions are excellent, yielding a clean image with
minimal PSF residuals. 
The POL*L images
are more affected by instrumental residuals due to 
optimization of the NIC2 coronagraphic optics for shorter wavelengths,
and to a better PSF template color match at 1.1~\micron.
These factors result in 2.0~\micron\ images dappled with faint residual speckles,
but that still clearly show the bright circumstellar nebulosity.

\section{Results}

A complex and asymmetric nebula surrounds AB Aurigae. 
The F110W image clearly shows the multiple spiral arms
previously observed, for instance by
\citet{Grady:1999p2806} and \citet{Fukagawa:2004p87}. 
The bright inner region of the disk extends out to
$\sim1.2\arcsec$, and is brightest to the south and southeast, as was
seen by Fukagawa et al. At 2.0~\micron\, in total intensity (Stokes $I$) the
overall surface brightness distribution is similar 
to that seen at 1~\micron, albeit at lower
angular resolution and contrast. 

\begin{figure*}[p]
  \includegraphics[width=7in]{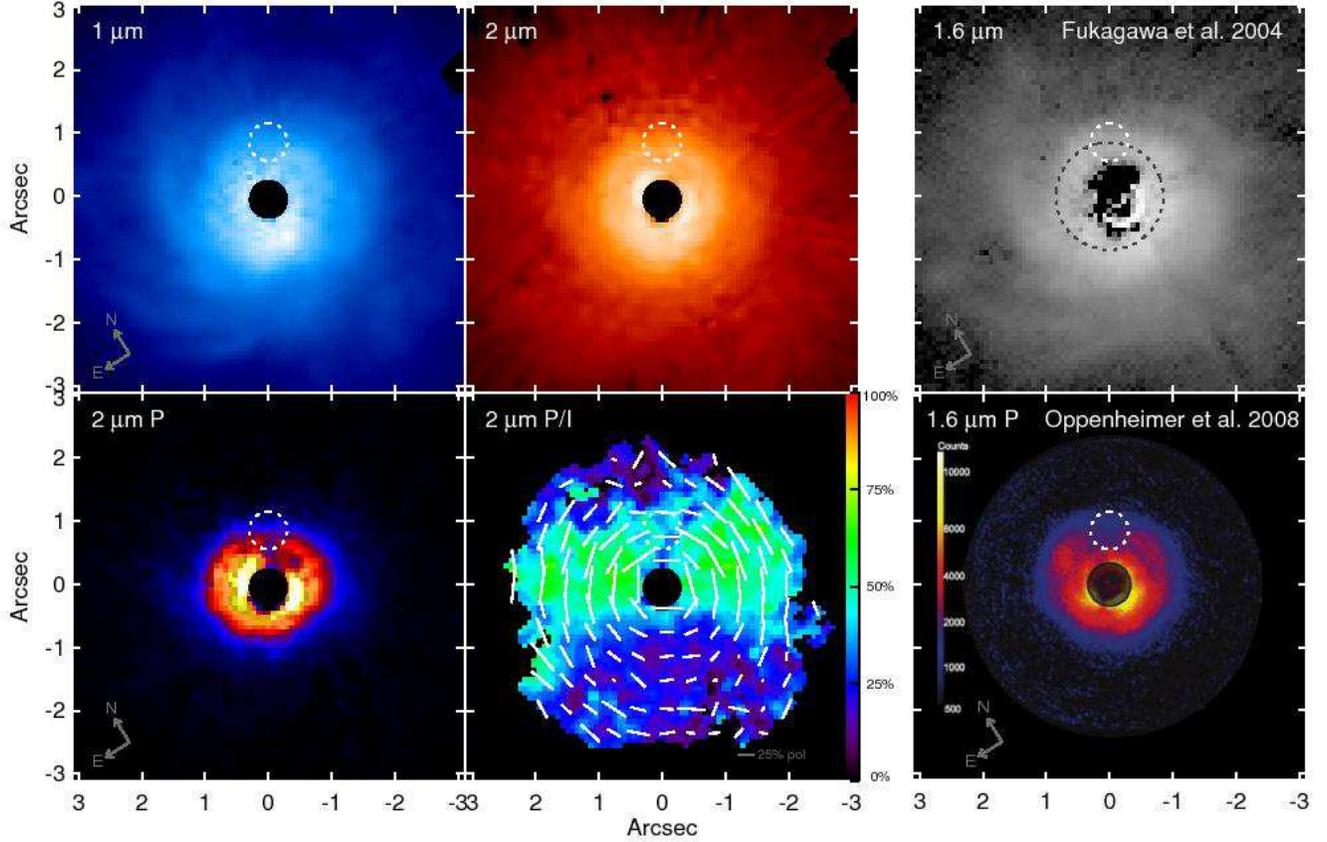}
  \caption{ Near-IR imaging and polarimetry of AB Aur. \textit{Left
  four panels: our new NICMOS observations}. These panels display
  the 1.1 and 2.0~\micron\ total intensities (using arcsinh stretches),
  2.0~\micron\ polarized intensity (using a log stretch matching that
  used by Opp08), and polarization fraction (shown using a linear
  stretch indicated by the inset color bar, and with vectors showing the polarization angles). All images have been
  rotated to align the disk's minor axis at PA=$328\degr$ vertical.
  \textit{Right panels:} $H$ band total intensity and polarized
  intensity from previous works, for comparison. \\
  Our 2~\micron\
  polarized intensity observations reproduce the appearance seen by
  Opp08 very closely, but our $P/I$ image shows that
  the ``depleted region'', indicated with a white dashed circle in all
  panels, is in fact a region of lower polarization
  fraction, not lower total intensity. None of the total intensity
  images show any indication of an open
  region at that location in the disk (though we caution that  
 in the $H$ band image from Fukagawa et al., the region of interest straddles the radius, shown in black, inside of which they state PSF subtraction artifacts rendered their
 data not photometrically reliable.)
  \label{heximages}
 }

\end{figure*}

The $2.0~\micron$ polarized
intensity image (lower left panel of Figure \ref{heximages})
reproduces with high fidelity
the polarized intensity pattern from Opp08. Both images show the region of lower polarized intensity at PA=333\degr\ 
between two brighter ``shoulders'' on either side,
the brightness enhancement southwest of the star, and even the scalloped, almost-concave
southern edge of the bright polarized region. Comparing the NICMOS and AEOS datasets, the NICMOS image  is more sensitive  and
traces polarized
light further from the star ($\sim7\arcsec$ vs.
$\sim1.2\arcsec$), while the
AEOS image has slightly better angular resolution, and better speckle rejection due to the
simultaneous differential technique \citep{Kuhn:2001p190,Perrin:2008p2351}.
Yet the most significant advantage of our NICMOS observations is
that they allow direct measurement the polarization fraction as 
the ratio $P/I$ (Figure \ref{heximages}, bottom center).

The percentage polarization revealed this way varies strongly 
around AB Aur: The disk's southeastern half is much less
polarized than the opposite side, with polarizations of $\sim$
$25\pm5\%$ and $45\pm12\%$, respectively. The two bright ``shoulders''
are seen to be regions of maximum polarization ($\sim55\%$),
separated by the lower polarization ($\sim40\%$) region corresponding to the ``depleted region'' from Opp08. 
The symmetry axis
of the overall polarization pattern is $328\degr\pm3\degr$, aligned with both
the ``gap'', and with the inferred rotation axis of
the disk \citep{Corder:2005p2693,Pietu:2005p1365}.  
The claimed ``gap'' in polarized intensity is now seen to be a
region of \textit{lower polarization fraction}, and \textit{not} a region of decreased total
disk-scattered light. There is no significant decrease in surface brightness in any of the total intensity images at this
location (see the white circles in Figure \ref{heximages}).

\begin{figure*}[p]
    \includegraphics[height=2.8in]{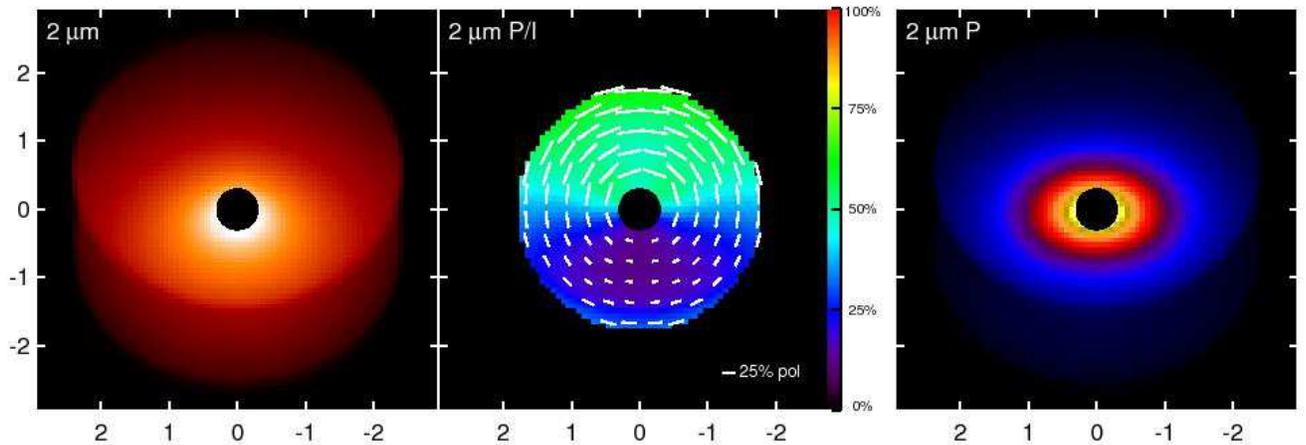}
  \vspace{-0.3in}
  \caption{ Our best-fit MCFOST simulated images of AB Aur, for an inclination of
  $35\degr$ and using a population of astrosilicate grains with 60\% porosity,
  $a_{max}=1~\micron$ and $dN/da\propto{a^{-3/5}}$. The three panels show 
  $2~\micron$ total intensity, polarization fraction, and polarized
  intensity, using the same display scales as the corresponding  panels in figure
  \ref{heximages}. See \S \ref{sect_model}.
  \label{MCFOST_im}
  }
\end{figure*}

\begin{figure}
  \includegraphics[width=3.5in]{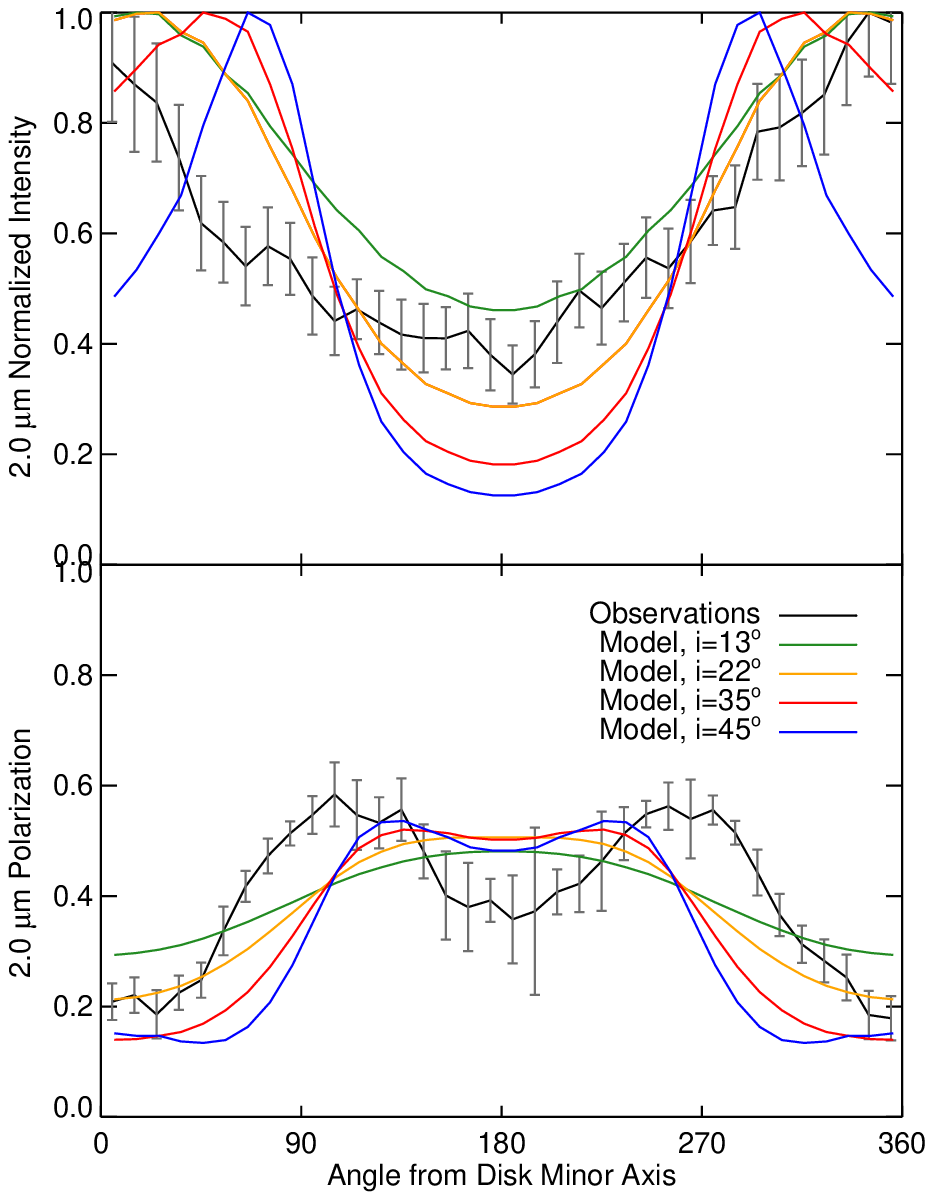}
  \caption{ 
  \label{MCFOST_profs}
    Observed and modeled azimuthal profiles for 2.0 $\micron$
    polarization and total intensity, measured in an annulus from 0.7--1.5$\arcsec$. 
    Azimuthal angles on the horizontal axis are measured relative to the minor axis pointing toward us.
    For both data and models, the near side is brightest in total intensity, but the far side
    is much more polarized. The models shown are for the same best fit
    grain population as in Figure 3.  For these dust parameters, the
    double-peaked pattern in polarized intensity is best fit at
    inclinations $\gtrsim35\degr$, but total intensity and the overall
    polarization level is best fit 
    at inclinations $\sim22\degr$. We conclude that our
    model's 60\% porous silicate spherical grains are only an
    approximate fit for the
    dust around AB Aur. 
    }
\end{figure}

\section{A Model for Scattering from AB Aur}

\label{sect_model}

The alignment of the polarization pattern with the disk's inclined rotational
axis provides a convincing indication that scattering geometry
and not disk substructure predominantly causes the observed appearance in polarized light. 
As shown in Fig.~1, the polarization of dust-scattered light is
maximized for scattering angles near or just above 90\degr, and then decreases for scattering angles closer to 180$\degr$.  The higher polarization of the 
northwestern half of the disk indicates that side is tilted away from us; the most distant part of the disk, where the light is most
strongly back-scattered, has lower polarization without any need to invoke clearing.

To demonstrate that this scenario accounts for the observed polarization
around AB Aur, we produced model images using the MCFOST
Monte Carlo radiative transfer code \citep{Pinte:2006p2701}.  
We concentrated on fitting the 2.0~\micron\ polarization by varying dust properties, 
holding most disk parameters (e.g. radii, scale height, etc.) fixed based on previous 
modeling \citep[e.g.][and references therein]{Tannirkulam:2008p2675}.
The inclination of AB Aur's disk is not precisely known: Using CO
kinematics, some authors have found inclinations as high as 33-40$\degr$
\citep{Pietu:2005p1365,Lin:2006p2790}, but others favor 
21$\degr$ or less \citep{Corder:2005p2693}. Lower inclinations are
also supported by the observed very low column
density of hydrogen and high Ly $\alpha$ wind velocity
\citep{Roberge:2001p2821}, and by near-infrared interferometry
\citep{Eisner:2004p249}. Due to the uncertainty, we allowed the
inclination to vary as a free parameter. 

We evaluated models across a grid in parameter space 
for plausible dust populations (see Table \ref{grid_table}). 
For each model, we computed the SED and 2
$\micron$ total and polarized images at a
range of inclinations. Images were convolved with a 0.22\arcsec\
Gaussian to match the data's resolution.
To find the best fit, we first discarded models 
which did not provide an acceptable match to AB Aur's SED, then
calculated each model's azimuthal polarization profile in an
0.7--1.5$\arcsec$ annulus, and computed the $\chi^2$ statistic
relative to the observed 2.0 $\micron$ polarization profile. 

The best fit model uses silicate grains with 60\% porosity, 
maximum grain size 1~\micron\ and a size distribution with power law index $= -3.5$.
See Figures \ref{MCFOST_im} and \ref{MCFOST_profs}. These parameters
are similar to those inferred for solar system cometary dust grains 
\citep{2000Icar..148..526P,Shen:2009p2830}.

We found that inclinations of 22-35\degr\ best fit the scattered light around AB
Aur, consistent with the range inferred from CO velocities. 
With the above dust parameters, inclinations of $\geq 35\degr$ better reproduce the
polarization drop from backscattering. 
Lower inclinations $\leq 22\degr$ improve the fit to the average polarization of the near and far sides, 
but do not show as 
strong a decrease in polarization on the far side as is observed.
Lower inclinations also better 
simultaneously fit the total intensity profile as well as
polarized intensity (See Figure \ref{MCFOST_profs}).

These slight discrepancies are most likely due to oversimplifications
in our dust model: 
MCFOST assumes simple Mie scattering from spherical dust grains, 
an imperfect approximation for actual circumstellar dust, which
is believed to consist of irregular fractal aggregates
\citep[][and references therein]{Dominik:2007p2624}. 
Realistic scattering properties for fractal grains can be calculated through 
more computationally intensive techniques such as the discrete dipole
model
\citep[e.g.][]{2000Icar..148..526P,2008MNRAS.390.1195D,Shen:2009p2830}. 
Aggregate grains seem promising candidates to provide a 
better fit for AB Aur. 

Specifically, discrete dipole calculations in
some cases predict polarizations that peak
at lower scattering angles (90\degr\ or below) and decrease more strongly toward
180\degr\ compared with Mie results (see for instance Figure 8 of Shen et al.\ 2009). That shift toward higher polarization at lower scattering
angles would broaden the azimuthal polarization profile
and could deepen a dip from backscattering, two changes that would
improve our fit to AB Aur. (In Figure 4, the visible offset between
the observed profile's polarization peaks, near 90 and 270\degr, and
the model profiles' peaks, near 135 and 225\degr, directly shows the
need for dust grains whose polarization maximum occurs at lower scattering
angle.)  Improving disk models to use more sophisticated dust
treatments during radiative transfer is a logical next step.

\begin{deluxetable}{ll}
\tablecolumns{2}
\tablecaption{\label{grid_table}Range of parameters explored in model grid}
\tablehead{ Parameter & Values }
\startdata
Inclination & 13--89\degr, 20 steps evenly spaced in $\cos i$ \\
Dust composition& astrosilicates\tablenotemark{a},
olivine\tablenotemark{b}, \\
	    &  ISM silicates+carbon mixture\tablenotemark{c}\\
Grain porosity & 0, 0.2, 0.4, 0.6, 0.9 \\
Grain maximum radius & 1, 20, 200 \\
Grain size power law & -2.5, -3.5, -4.5 \\
\enddata
\tablecomments{
Fixed model parameters: Stellar $T_{eff}=9772$ K, $L_*=47 L_\odot$, dust
mass $10^{-4} M_\odot$, disk $R_{inner}=0.2$~AU, $R_{outer}=350$~AU, scale height $h=14$ AU at $R=100$ AU, disk flaring $h\propto R^{1.3}$. 
See \citet{Tannirkulam:2008p2675} and references therein.
}
\tablenotetext{a}{\citet{1984ApJ...285...89D,2001ApJ...548..296W}}
\tablenotetext{b}{\citet{1995AnA...300..503D}}
\tablenotetext{c}{\citet{1989ApJ...341..808M}}
\end{deluxetable}

\section{Discussion:}

\label{discussion}

\subsection{Structure in AB Aurigae's Disk}

The pattern of polarization resolved around AB Aur 
indicates unambiguously that the 
spatial variation of polarized light there is due primarily to the geometry of
scattering from the inclined disk's surface. 
These data do not support the hypothesis of significant clearing in the disk near PA=333\degr.

Independent of but simultaneously with this study, a numerical 
investigation of protoplanet shadows in disks reached a similar conclusion,
finding the observed morphology around AB Aur inconsistent with the presence of any protoplanet above 0.3 Jupiter
masses (Jang-Condell and Kuchner, submitted).

While the previously-claimed gap does not appear to be present, there \textit{is} real
structure within AB Aur's disk which is visible in scattered light,
including polarization.
Many of the asymmetric spots
seen in the 2.0~\micron\ polarization fraction image coincide with the spiral arms as seen in 
the 1.1~\micron\ image. For instance, the northeasternmost region of high
polarization, near (1.7,~0) in the coordinate system of Figure 2,
is precisely aligned with the brightest spiral arm. Overplotting or blinking these images shows several
such alignments.  Perturbations in the disk's surface, from either 
localized changes in scale height or warps in midplane location, 
could change the scattering geometry at the optical depth $\tau=1$
surface to produce the observed minor variations in
polarization. For optically thick disks, surface features seen in scattered
light do not necessarily correlate with conditions at the midplane
\citep{JangCondell:2007p447}, but in the case of AB Aur there is
direct evidence from CO emission that the bulk of the gas deviates
from pure Keplerian rotation \citep{Lin:2006p2790}.
These perturbations and/or warps might contribute to the difficulty in
firmly establishing the system's inclination. But this just raises
the inevitable next question: what causes those fluctuations?

Though we find no direct evidence for any disk gap due to a protoplanet, there still remains a 
case for the presence of a companion somewhere around AB Aur: 
Both the strong
spiral structure seen in dust-scattered light and the non-Keplerian dynamics revealed by gas emission lines argue for the existence of a planetary-mass body
perturbing the disk. Other perturbation mechanisms, such as gravitational instabilities in the disk or
a stellar-mass companion, seem ruled out observationally 
\citep{Pietu:2005p1365,Lin:2006p2790}.  The detection of this
companion thus awaits future improvements in high-contrast imaging to
reveal.

\subsection{Interpreting Features in Imaging Polarimetry}

Differential polarimetry, as used by Opp08 and others, is a proven technique
for observing circumstellar dust at
high contrast from the ground with
adaptive optics (AO).
Such observations will become increasingly common 
with upcoming extreme AO systems, such as GPI \citep{2006SPIE.6272E..18M} and SPHERE \citep{2006Msngr.125...29B}. 
For many disks, these instruments will provide high contrast images only in polarized light. 

We have shown here that care is required when interpreting such data.
Variation of polarization with scattering angle cannot be neglected,
even for low inclination targets like AB Aur. 
Yet AO differential polarimetry can still
yield precise measurements of disk structure and dust properties,
provided the degeneracy between polarization fraction and intensity can be broken. This may be possible
through multiwavelength observations, which are differently sensitive
to scattering geometry \citep{Watson:2007p2642}, or in combination
with independent constraints on geometry such as from CO emission.

For brighter disks, extreme AO should allow
accurate measurement of polarization 
fraction from the ground, but for now, 
HST NICMOS offers the highest polarimetric precision for 
measurements of disk-scattered light.

\label{concl_sect}

\acknowledgements

Based on observations made with the NASA/ESA Hubble Space Telescope.
Support for program GO-11155
was provided by NASA through a grant 
from the
Space Telescope Science Institute, which is operated by the
Association of Universities for Research in
Astronomy, Inc., under NASA contract NAS 5-26555. 
MDP is supported by an NSF
Astronomy \& Astrophysics Postdoctoral Fellowship.
GS and DH were also supported by programs HST/GO 10847 and 10852.
MDP thanks Ben Oppenheimer and Misato Fukagawa for 
discussions, and for sharing their data in FITS format.

\end{document}